# Ti substituted $La_{0.67}Ca_{0.33}MnO_3$ ortho-perovskites:
## Dominant role of local structure on the electrical transport and magnetic properties

L. Seetha Lakshmi,[*,@] V. Sridharan,[*] R. Rawat,[§] and V. S. Sastry[*]

[*]XS & CGS, Materials Science Division, Indira Gandhi Centre For Atomic Research, Kalpakkam, 603102 Tamil Nadu, India
[§]UGC-DAE Consortium for Scientific Research, Khandwa Road, Indore 452 017, Madhya Pradesh, India

(18th December 05)

**Abstract**

We report the effects of local structure on the electrical transport and magnetic properties of $La_{0.67}Ca_{0.33}Mn_{1-x}Ti_xO_3$ ($0 \leq x \leq 0.10$) system. Linear increase in the lattice parameters, consequent expansion of unit cell and a monotonic decrease in the relative $Mn^{4+}$ concentration with x suggest that $Ti^{4+}$ predominantly replaces $Mn^{4+}$ in $La_{0.67}Ca_{0.33}Mn_{1-x}Ti_xO_3$ ($x \leq 0.10$). The ferromagnetic-metallic ground state modifies to a glassy insulator for $x \geq 0.05$. No field induced metallic state could be discerned for x = 0.10 even at a field of 8 T and at as low a temperature as 4.2 K. Ti substitution significantly enhances the colossal magnetoresistance effect. Both the metal to insulator transition temperature and Curie temperature ($T_c$) decrease at a rate of ~ 26 K/at.% up to x = 0.05. $T_c$ levels off for higher compositions. Modification of the major carrier concentration (decreased $Mn^{4+}$ concentration) seems to be insufficient to account for the observed reduction in the transition temperatures. This in turn emphasizes the significance of local structural effects: systematic elongation of the Mn-O bond lengths and decrease of Mn-O-Mn angles leading to strong suppression of itinerant ferromagnetism and metallicity of the compounds. The additional features in the ac susceptibility, viz., a broad shoulder just below $T_c$ followed by a sharp decrease in the susceptibility signal at low temperatures and a non-closure of MR at zero field as high as ~ 75 % at 5 K indicate a frustrated magnetic ground state for x = 0.07. Based on the inter-comparison of the structural, electrical transport and magnetic properties of the Mn site substituted $La_{0.67}Ca_{0.33}MnO_3$ with iso-valent diamagnetic and paramagnetic ions, we argue that local structural effects have a decisive role to play, compared to the local spin coupling effects, in the ferromagnetic-metallic ground state of the CMR manganites.

**Introduction**

The hole doped mixed-valent $La_{0.67}Ca_{0.33}MnO_3$ ortho-perovskite undergoes a concomitant Curie transition and a metal insulator transition (MIT) at ~ 270 K with the largest reported colossal magnetoresistance (CMR).[1] The basic mechanism which couples the charge and the spin degrees of freedom in the low temperature ferromagnetic-metallic (FM-M) state is the double-exchange (DE), initially proposed by Zener.[2] Under this scheme the electron transfer integral t, representing the kinetic energy of the electron, is modified as $t = t_o \cos(\theta/2)$, where $t_o$ is the bare electron transfer integral and θ is the angle between two neighbouring Mn spins. The magnitude of $t_o$ essentially depends on the structural effects and θ is sensitive to the nature of the local spin coupling of the paramagnetic ions. The validity of the DE model was called into question by Millis et.al,[3,4] since it fails to account for many of the observed properties, including the insulating behaviour in the paramagnetic regime above $T_c$ and the CMR phenomena. Subsequent studies suggested that the strong electron-phonon coupling due to the presence of Jahn-Teller (JT) $Mn^{3+}$ ion is required to understand the rich variety of phenomena found in these compounds.[4,5,6,7,8] Among the other important issues, numerous theoretical and experimental investigations in recent years have established the key role of intrinsic inhomogeneites and mixed phase tendencies on the observed colossal effects and the complex ground state properties of the compounds.[9] Thus, the studies addressing the disorder effects, experimentally achieved by a suitable substitution at La/Ca[10,11] and Mn sites, have revived considerable interest. In this scenario, Mn site substitution studies deserve special attention, since it offers a useful handle to manipulate the charge, spin, lattice and orbital degrees of freedom.

Previous studies of Mn-site substitution on the $La_{0.67}Ca_{0.33}MnO_3$ system indicate that the metal to insulator (MIT) transition temperature ($T_{MI}$) and Curie temperature ($T_c$) are suppressed but to different extent for different substituents.[12,13,14,15,16,17,18] The suppression in transition temperatures and the modification of the ground state properties were broadly ascribed to the 'weakening' of the DE interaction strength. However, the factors leading to the extent of variation in the suppression of the transition temperatures and ground state properties remain unexplained and this needs a further systematic investigation. In our earlier work on Mn-site substitution with para and diamagnetic ions, two factors: local structural and local spin coupling effects have been found to affect the transition temperatures.[19,20] In this paper, we report the importance of local structural effects on the ground state properties through Ti doping in the $La_{0.67}Ca_{0.33}MnO_3$ compound. Tetravalent Ti exhibits $d^0$ electronic configuration and it is expected not to introduce additional effects due to magnetic coupling. Our study show that reduction of majority carrier concentration cannot account for the observed strong reduction of the transition temperatures of the compounds. This further strengthens our argument that local structural effects have a decisive role compared to local spin coupling effects on the transition temperatures and the FM-M ground state of the CMR manganites.

---

[@] *Corresponding Author - email: slaxmi73@gmail.com (L. Seetha Lakshmi)*

**Experiment**

Bulk polycrystalline $La_{0.67}Ca_{0.33}Mn_{1-x}Ti_xO_3$ ($0 \leq x \leq 0.10$) compounds were prepared by standard solid state reaction using stoichiometric amount of precursors, $La_2O_3$ (INDIAN-RARE EARTHS), $CaCO_3$ (CERAC), $MnO_2$ (CERAC) and $TiO_2$ (CERAC), all with purity better than 99.995 %. Care was taken to remove the moisture in $La_2O_3$ before weighing by preheating at 800 ºC for 24 hours. High purity acetone was used as a wetting medium to assist the mixing process of the precursors. The homogenized, dried powder was calcined at 980 ºC for 24 hours and subsequently ground and compacted into pellets of 15 mm diameter. The pellets were heat treated in the temperature range 1200 to 1425 ºC for 36 hours in flowing oxygen with four intermediate grindings followed by pelletization. Pellets were further ground into very fine powder and compacted into pellets of 15 mm diameter and ~ 1.5 mm thickness. Polyvinyl acetate (PVA) (2 wt. %) solution was used as a binding material to improve the inter-grain connectivity. The final sintering of pellets, all in a single batch, was carried out at 1525 ºC for 36 hours in flowing oxygen. Sufficient care was taken at the different stages of the sample preparation, so that the preparation history is the same for all samples prepared. Typical density of the sintered pellets, determined by modified Archimedes method using ethanol as liquid, was ~ 92-95 % of the theoretical density and no systematic variations were observed with Ti composition.

The room temperature powder X-ray diffraction (XRD) pattern in Bragg-Brentano para-focusing geometry with high statistics (~ $10^5$ counts over a dwell time of 25 sec at the 100 % peak) was recorded using $Cu_{K\alpha}$ radiation (STOE, Germany). Si powder (NIST SRM-640b) was used as an external standard for 2θ (zero) correction before recording each new set of XRD pattern. The structural refinement was carried out using GSAS Rietveld refinement program.[21] SPUDS program was used for generating the starting model for the Rietveld analysis.[22] For all compounds, the site occupancy was constrained to be consistent with the nominal stoichiometry of the compound. The atomic co-ordinates and isotropic thermal parameters were constrained to be equal for the different atomic species having the same point symmetry. They were allowed to vary independently for different compositions. The resistivity measurements (ρ(T)) were performed on rectangular bar shaped pellets in linear four-probe geometry using a $^3$He flow cryostat (OXFORD INSTRUMENTS, UK). A superconducting magnet was employed for steady magnetic fields up to 8 T with the magnetic field parallel to current. The metal-insulator transition (MIT) temperature ($T_{MI}$) is the temperature corresponding to the maximum in the ρ(T) curve. The magnetoresistance (MR) is defined by $MR(\%) = \left((\rho_{H \neq 0} - \rho_{H=0})/\rho_{H=0}\right) \times 100$ where $\rho_{H \neq 0}$ and $\rho_{H=0}$ are the resistivity in the presence and absence of the magnetic field respectively. To study the MR as a function of field, samples were cooled in zero field from room temperature to 5 K. The magnetic field was swept from 0 T to +5 T and then back to 0 T and resistance was recorded along the field sweep. Temperature variation of ac susceptibility with ac probe field (h) of 0.25 Oe and an excitation frequency (f) of 941 Hz was measured using a home made susceptometer. The in-phase component of ac susceptibility (χ′) alone is considered for further discussion. The Curie temperature ($T_c$), corresponding to the onset temperature of the susceptibility signal, was estimated by the tangent method.

**Results and Discussion**

High statistics RT powder XRD patterns (**Fig.1**) indicate that all the compounds are iso-structural with orthorhombic Pnma symmetry (Space Group No. 62).The Rietveld refinement yielded excellent agreement between the calculated and observed positions of the diffraction profiles. No extra reflections other than those allowed by the space group and symmetry conditions were observed indicating that $La_{0.67}Ca_{0.33}Mn_{1-x}Ti_xO_3$ ($0 \leq x \leq 0.10$) compounds are single-phase in nature. As a representative of the series, the Rietveld refinement spectra of x = 0 and 0.10 are shown in **Fig.2(a) and (b)** respectively. Structural parameters deduced from the Rietveld analysis are summarized in **Table 1**. In **Table 1**, are also reported the reliability factors ($R_{wp}$ and $R_p$) and the goodness of fit (S). Comparatively higher values of the R-factors might be due to the fact that refinement studies were performed without background subtraction. Typical S value of 1.4 indicates that the quality of fit is good.

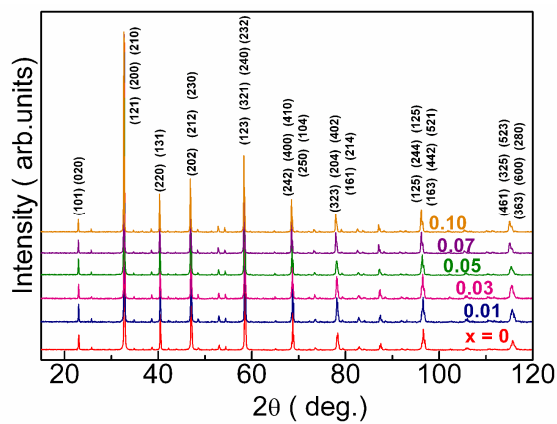

**Fig. 1**: High statistics room temperature powder XRD patterns of $La_{0.67}Ca_{0.33}Mn_{1-x}Ti_xO_3$ ($0 \leq x \leq 0.10$) compounds. Miller indices of the major Bragg reflections are also indicated.

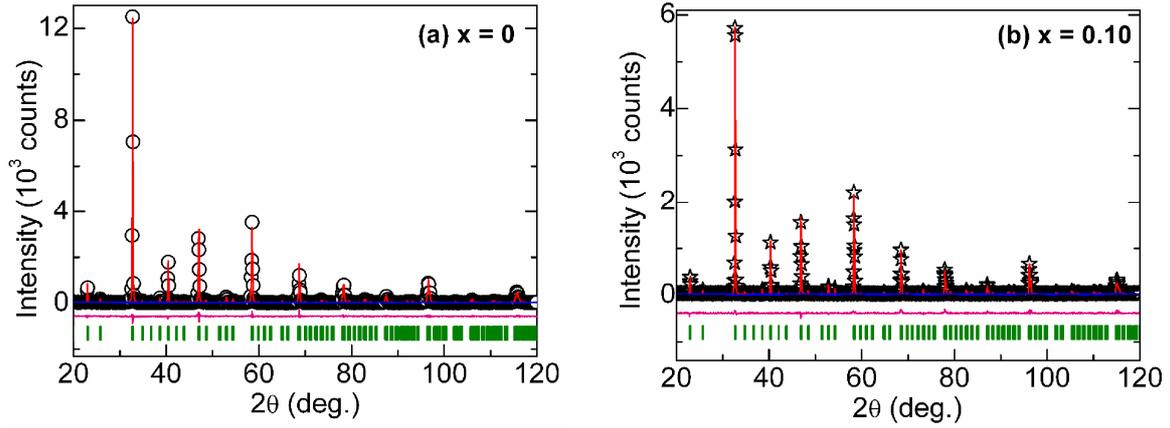

**Fig. 2**: Observed (open symbols) and calculated (solid lines) XRD patterns for $La_{0.67}Ca_{0.33}Mn_{1-x}Ti_xO_3$ compounds for x = 0 (bottom panel) and x = 0.10 (top panel). Difference between the observed and the calculated intensities are shown at the bottom of the figure. Positions for the calculated Bragg-reflected positions are marked by the vertical bars.

**Table 1**: The refined lattice parameters (*a, b* and *c*) (in Å), unit cell volume (v) (in Å$^3$), estimated Mn-O bond lengths ($d_{Mn-O1}$ and $d_{Mn-O2}$) (in Å), average Mn-O bond length ($<d_{Mn-O}>$) (in Å), Mn-O-Mn bond angles (Mn-O1-Mn and Mn-O2-Mn) (in deg.), average Mn-O-Mn bond angle ($<Mn-O-Mn>$) (in deg.) from the fractional co-ordinates, relative concentration of $Mn^{4+}$ ion (in %), reliability factors: $R_{wp}$ and $R_p$ (in %) and goodness of fit (S) from the Rietveld analysis of the RT powder XRD patterns of $La_{0.67}Ca_{0.33}Mn_{1-x}Ti_xO_3$ ($0 \leq x \leq 0.10$) compounds. Numbers in parenthesis indicate the standard deviation in the last digit. The atomic positions in the orthorhombic Pnma symmetry are 4c (x, 0.25, z) for La/Ca , 4b (0.5, 0, 0) for Mn/Ti , 4c (x, 0.25, z) for O1 and 8d (x, y, z) for O2. Here O1 is the apical oxygen and O2 is the equatorial oxygen in the plane of $MnO_6$ octahedra.

| x | 0 | 0.01 | 0.03 | 0.05 | 0.07 | 0.10 |
|---|---|---|---|---|---|---|
| a | 5.4574(1) | 5.4604(2) | 5.4637(1) | 5.4684(0) | 5.4734(2) | 5.4790(2) |
| b | 7.7088(1) | 7.7126(1) | 7.7191(1) | 7.7259(1) | 7.7339(1) | 7.7397(1) |
| c | 5.4730(0) | 5.4747(1) | 5.4765(1) | 5.4790(0) | 5.4821(1) | 5.4844(2) |
| v | 230.255(4) | 230.562(5) | 230.968(8) | 231.485(4) | 232.061(6) | 232.570(6) |
| $d_{Mn-O1}$ | 1.95344(4) | 1.95452(4) | 1.95689(2) | 1.96075(3) | 1.96385(4) | 1.96679(4) |
| $d_{Mn-O2}$ | 1.96652(3) | 1.96763(2) | 1.96975(1) | 1.97096(4) | 1.97194(1) | 1.97293(0) |
| $<d_{Mn-O}>$ | 1.95982(4) | 1.96108(3) | 1.96332(2) | 1.96586(4) | 1.96790(3) | 1.96986(4) |
| Mn-O1-Mn | 161.202(1) | 160.950(3) | 160.500(0) | 159.511(1) | 158.380(2) | 157.711(3) |
| Mn-O2-Mn | 161.926(0) | 161.651(0) | 161.343(1) | 160.259(1) | 159.294(3) | 158.832(1) |
| $<Mn-O-Mn>$ | 161.564(1) | 161.300(3) | 160.921(1) | 159.885(1) | 158.832(3) | 158.206(2) |
| $Mn^{4+}$ | 33 | 32.2 | 30.1 | 27.9 | 26.1 | 22.8 |
| $R_{wp}$ | 11.55 | 16.98 | 17.08 | 11.10 | 16.05 | 15.36 |
| $R_p$ | 8.50 | 12.05 | 12.71 | 8.07 | 11.74 | 11.23 |
| S | 1.35 | 1.46 | 1.42 | 1.35 | 1.37 | 1.40 |

The lattice parameters (a, b and c) (**Fig.3**) and thus the unit cell volume (v) (**inset of Fig.3**) increase linearly with Ti concentration. The Rietveld refinement studies show that $Ti^{4+}$ systematically decreases the $Mn^{4+}$ concentration (**Table 1**) according to $La^{3+}_{0.67}Ca^{2+}_{0.33}Mn^{3+}_{0.67}Mn^{4+}_{(0.33-x)}Ti^{4+}_xO^{2-}_3$. This is in general agreement with the previous report.[23] The ionic radii of $Mn^{3+}$, $Mn^{4+}$ and $Ti^{4+}$ ions for a six-fold co-ordination are 0.645, 0.53 and 0.605 Å respectively.[24] Thus, the observed unit cell expansion is due to an increase in the average Mn radius from 0.6071 Å for x = 0 to 0.6145 Å for x = 0.10. However, a much larger fractional increase in a, b, c and v of 0.4, 0.4, 0.2 and 1.0 % respectively are observed for x = 0.10 in our case, nearly three times larger than reported by Lu et.al [25] is noteworthy. The observed unit cell expansion together with the systematic reduction in the Mn$^{4+}$ concentration with x indicates that Ti$^{4+}$ predominantly substitutes Mn$^{4+}$ ion. Furthermore, Ti substitution leads to large local structural effects involving an almost linear elongation of the average Mn-O bond length ($<d_{Mn-O}>$) and a reduction in the average Mn-O-Mn bond angle ($<Mn-O-Mn>$) (**Fig.4**). This is expected to have a dominant effect on the electrical transport and the magnetic properties as discussed in the following.

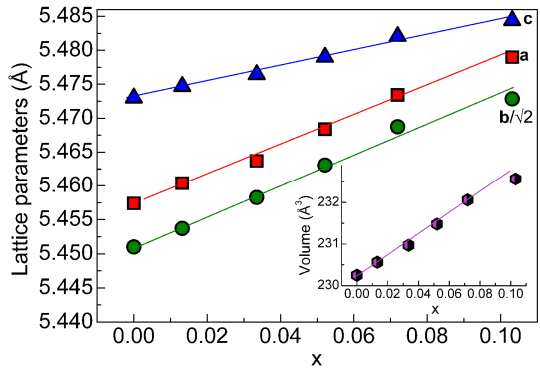

Fig. 3: The composition dependence of (a) lattice parameters, $a$ ($\square$), $b$ ($\bigcirc$) and $c$ ($\triangle$) (in Å), unit cell volume, $V$ ($\bigcirc$) (in Å$^3$) (inset of the figure) of La$_{0.67}$Ca$_{0.33}$Mn$_{1-x}$Ti$_x$O$_3$ ($0 \leq x \leq 0.10$) compounds. The dotted line is only a guide to eye. Error bars are smaller than symbols.

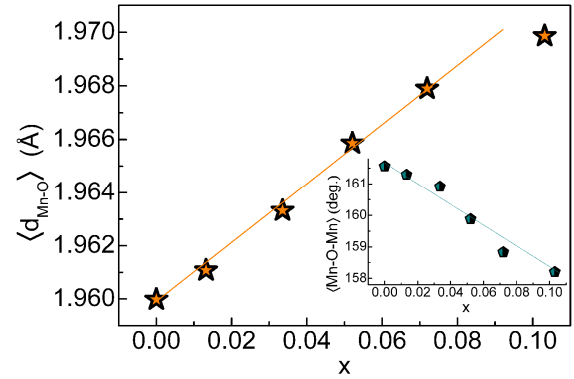

Fig. 4: The composition dependence of average Mn-O bondlength ($<d_{Mn-O}>$) ($\star$) (in Å) and average Mn-O-Mn bond angle ($<Mn-O-Mn>$) ($\bullet$) (in deg.) (inset of the figure) of La$_{0.67}$Ca$_{0.33}$Mn$_{1-x}$Ti$_x$O$_3$ ($0 \leq x \leq 0.10$) compounds. The dotted line is only a guide to eye. Error bars are smaller than symbols.

In **Fig.5**, the $\rho(T)$ curves in the absence of magnetic field are shown. All the compounds except x = 0.10 exhibit an MIT characterized by a maximum in $\rho(T)$ at T = T$_{MI}$ and the values are provided in **Table 2**. Though an MIT could be discerned in $\rho(T)$ for x = 0.07, $\rho$ is found to increase further after reaching a minimum value at ~ 30 K. Since our experimental set up limits the resistivity measurement to 10$^6$ $\Omega$.cm, no data could be collected below ~ 63 K for x = 0.10 (**Fig.5**). Nonetheless, the sample was further cooled down with frequent check on the value of the resistivity and such checks show that the resistivity even at 4.2 K is beyond the measurable range indicating the non-metallic nature of the compound. Residual resistivity ($\rho_0$) is found to increase gradually up to x = 0.05 and then shows a rapid rise of 6 orders of magnitude for x = 0.07

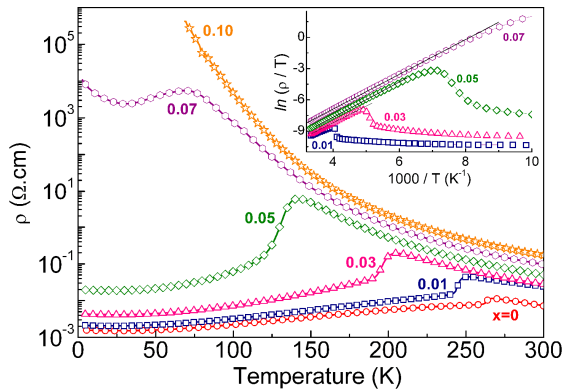

Fig. 5 : Temperature variation of resistivity ($\rho(T)$) under zero field of La$_{0.67}$Ca$_{0.33}$Mn$_{1-x}$Ti$_x$O$_3$ ($0 \leq x \leq 0.10$) compounds. Inset shows $\ln(\rho/T)$ versus $1/T$ curves of the compounds for x = 0.01, 0.03, 0.05 and 0.07.

(**Table 2**). As shown in the **inset of Fig.5**, we found that the resistivity in the high temperature paramagnetic-insulating regime for $0.01 \leq x \leq 0.07$ can be fitted to adiabatic small polaron hopping model,[26] $\rho = BT\exp(E_H/k_BT)$, where B $\left(=\dfrac{g_d k_B}{ne^2\alpha^2\upsilon_o}\right)$ is resistivity co-efficient, $E_H$ is the thermally activated hopping energy, $k_B$ is Boltzmann constant, $g_d$ is a geometrical factor (1 for simple cubic lattice and 2/3 for triangular lattice), n is the polaron concentration, $\alpha$ is the hopping distance and $\upsilon_O$ is the longitudinal optical phonon frequency. Due to the proximity of T$_{MI}$ to room temperature, such a fitting could not be done for the undoped (x = 0) compound. For x = 0.10, the resistivity data does not fit to the small polaron model well. But the low temperature resistivity could be fitted to variable range hopping model,[27] $\rho \approx \exp(T_0/T)^{1/4}$. It is seen that $E_H$ increases with x (**Fig.6**) in general agreement with the previous report.[18] On the other hand, B shows a rather complex dependence on x (**Fig.6**), similar to what has been observed in La$_{0.67}$Ca$_{0.33}$Mn$_{1-x}$Ga$_x$O$_3$ ($x \leq 0.10$) compounds.[14] B decreases with increasing x ($x \leq 0.05$) and rises for x = 0.07 compound. Ti$^{4+}$ ion, due to its closed shell configuration, acts as a blocking site for the thermally activated conduction process. Thus the carriers have to hop over such barriers and as a consequence, the the average hopping distance ($\alpha$) increases. The decrease of B with increase in x for lower Ti concentration is mainly due to the increase of $\alpha$. However, a rise of B for x = 0.07 could not be explained with the available experimental data. On the other hand, in $La_{0.67}Ca_{0.33}Mn_{1-x}Ga_xO_3$ (x = 0.10) compound,[14] a similar trend is observed and is attributed to the significant decrease in polaron concentration (n). $Ga^{3+}$ is known to replace $Mn^{3+}$ ion and hence the $Ga^{3+}$ substitution is expected to decrease n. Rietveld refinement results (**Table 1**) as well as the previous report [23, 25] clearly indicate that $Ti^{4+}$ replaces $Mn^{4+}$ ion in $La_{0.67}Ca_{0.33}Mn_{1-x}Ti_xO_3$ (x ≤ 0.10) compounds. Hence, a reduction in n is not expected in the present case. This indicates that decrease in n could not be a major reason for a rise in B for x = 0.07.

**Table 2**: Metal-Insulator transition temperature ($T_{MI}$) (in K) in a magnetic field (H) of 0, 1 and 5 T, Curie temperature ($T_c$) (in K), residual resistivity ($\rho_o$) (in $\Omega$ cm), temperature corresponding to the maximum in magnetoresistance at 1 and 5 T ($T_{MR}$) (in K), the magnetoresistance at the maximum in MR(T) in 1 and 5 T ($MR_1$ and $MR_5$) (in %), Grain boundary contribution to MR estimated from LFMR (GBMR) (in %) of $La_{0.67}Ca_{0.33}Mn_{1-x}Ti_xO_3$ ($0 \leq x \leq 0.10$) compounds.

| | x | 0 | 0.01 | 0.03 | 0.05 | 0.07 | 0.10 |
|---|---|---|---|---|---|---|---|
| $T_{MI}$ | H = 0 | 267.3 | 251.2 | 204.8 | 141.9 | 69.6 | - |
| | 1 | 278.0 | 261.7 | 215.9 | 157.3 | 78.4 | - |
| | 5 | > 300 | > 300 | 255.3 | 196.8 | 85.7 | - |
| $T_c$ | | 264.4 | 246.6 | 199.5 | 133.5 | 113 | 98.3 |
| $\rho_o$ | | 0.001 | 0.002 | 0.004 | 0.020 | 8962.3 | - |
| $T_{MR}$ | H = 1 | 266.5 | 248.2 | 201.1 | 137.3 | 45.2 | - |
| | 5 | 266.5 | 249.6 | 204.7 | 140.2 | 54.8 | - |
| $MR_1$ | | 37.3 | 66.4 | 78.5 | 93.3 | 93.5 | - |
| $MR_5$ | | 58.7 | 80.5 | 91.9 | 99.1 | 99.4 | - |
| GBMR | | 23.4 | 21.3 | 22.0 | 22.1 | - | - |

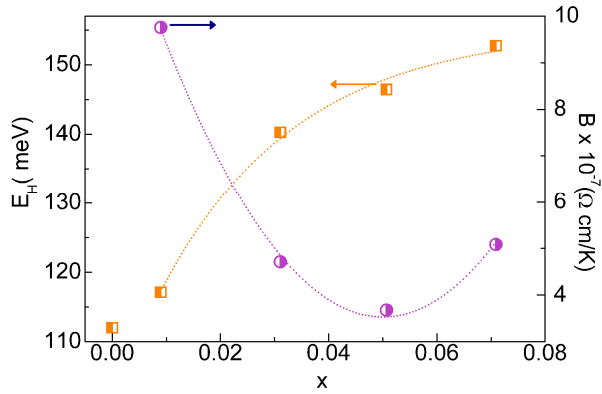

**Fig. 6**: Thermally activated hopping energy ($E_H$) (in meV) (□) and resistance co-efficient (B) (in $\Omega$ cm.K$^{-1}$) (○) as a function of x of $La_{0.67}Ca_{0.33}Mn_{1-x}Ti_xO_3$ ($0.01 \leq x \leq 0.07$) compounds. Continuous lines are a guide to eye and error bars are smaller than symbols.

All compounds, including the one not exhibiting an MIT, show a para to ferromagnetic transition at $T_c$, a quintessential feature of the CMR manganites (**Fig.7**). $T_c$, closely matching $T_{MI}$ within ± 5 K also decreases at the same rate (~ 26 K/at.%) up to x = 0.05 and then levels off for the higher composition (**Fig.8**). There exists a discernable difference of ~ 43 K between $T_c$ and $T_{MI}$ of x = 0.07 as shown. For x = 0.07, transition to metallic state occurs at temperature far below $T_c$, absence of any such perceptible anomaly in $\rho(T)$ near $T_c$ for x = 0.10 indicates the non-metallic nature of the ferromagnetic phase of this compound. Another striking feature to note for x = 0.07 is the broad shoulder in $\chi'(T)$ below $T_c$ and a sharp decrease of $\chi'$ at low temperatures. On the other hand, x = 0.10 shows a sharp cusp like feature just below $T_c$. These features clearly indicate the ground state is no longer a ferromagnetic metal. However, we shall not be concerned here with the nature of the modified magnetic ground state of these compounds.

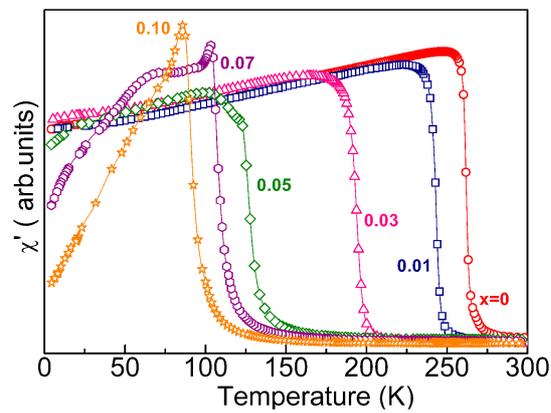

Fig.7 : The temperature variation of the in-phase component of ac susceptibility ($\chi'(T)$) of $La_{0.67}Ca_{0.33}Mn_{1-x}Ti_xO_3$ ($0 \leq x \leq 0.10$) compounds measured in an ac probe field of 0.25 Oe and an excitation frequency of 941 Hz.

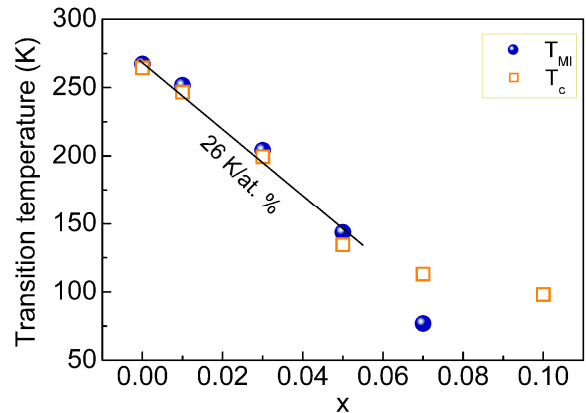

Fig. 8: Variation of metal-insulator transition temperature ($T_{MI}$) and Curie temperature ($T_c$) as a function of x of $La_{0.67}Ca_{0.33}Mn_{1-x}Ti_xO_3$ ($0 \leq x \leq 0.10$) compounds. Straight line is the best fit to the experimental data to estimate the rate of suppression in the transition temperatures.

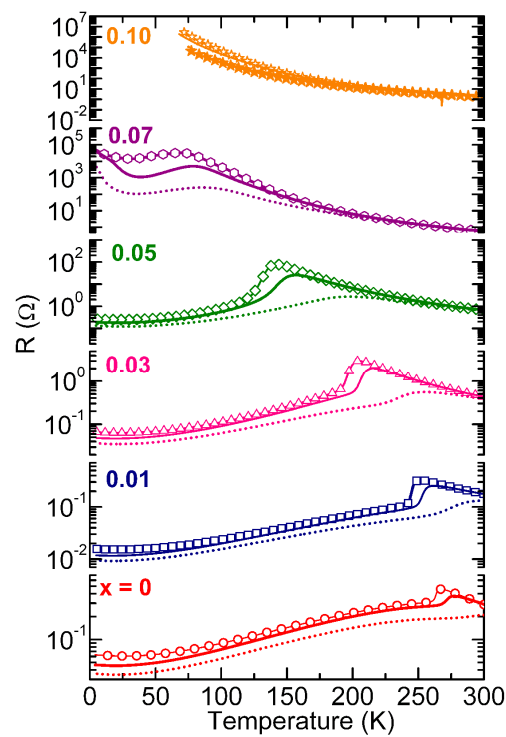

Fig. 9 : Temperature dependence of resistance (R(T)) of $La_{0.67}Ca_{0.33}Mn_{1-x}Ti_xO_3$ ($0 \leq x \leq 0.10$) compounds under a magnetic field of 0 (solid line), 1 (open symbols) and 5 T (dotted line). Closed symbols denote the R(T) curve of x = 0.10 in a magnetic field of 8 T.

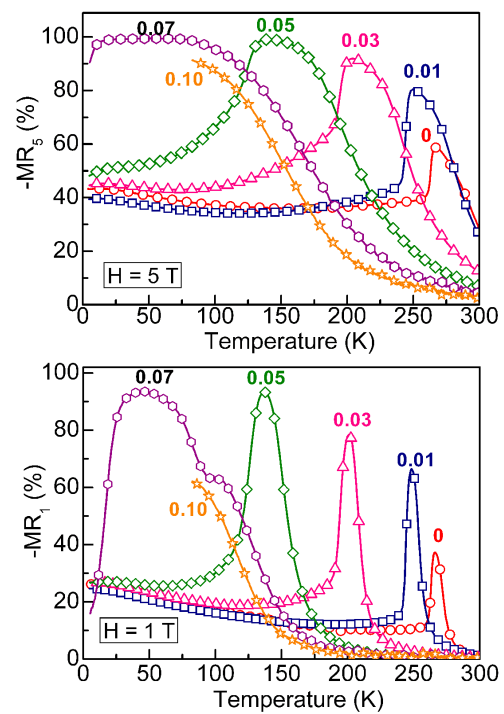

Fig.10 : Temperature dependence of magnetoresistance (MR(T)) of $La_{0.67}Ca_{0.33}Mn_{1-x}Ti_xO_3$ compounds ($0 \leq x \leq 0.10$) in a magnetic field of 1 T (bottom panel) and 5 T (top panel).

**Figure 9** presents the temperature dependence of resistance in a field of 1 and 5 T. For all compounds, except x = 0.10, the maximum is found to shift to higher temperature (**Table 2**) and broaden with an increase in H. A significant reduction in the resistance on the application of magnetic field is also observed over the entire temperature range of study. No field induced metallic state could be observed even at a field of 8 T and as a low temperature as 4.2 K, the resistance of the compound was beyond the measurable range of the instrument. For all compounds, except x = 0.10, temperature dependence of MR curves display a maximum at a temperature slightly below $T_{MI}$ (at H = 0) (**Fig. 10**). It is found that with increasing magnetic field, the MR maximum moves close to $T_{MI}$ in zero field. Furthermore, the magnitude of the MR maximum is significantly enhanced by Ti substitution for $x \leq 0.07$. Though the insulating Ti compound (x = 0.10) also exhibits an incipient peak like feature, the existence of peak could not be confirmed owing to non-availability of data. At 300 K, the MR(H) curves display a parabolic behaviour (**top panel of Fig.11**). For any given field strength, MR is found to increase with Ti concentration. On the other hand, for compounds with $x \leq 0.05$, MR(H) at 5 K exhibits two distinct regions (**bottom panel of fig.11**). A sharp linear fall at low fields (< 0.5 T) is attributed to spin polarized tunnelling through grain boundaries [28] and is shown to be a characteristic feature of polycrystalline manganites. When the field exceeds 0.5 T, MR of the compounds decreases with H. In the phase separation model, this could be understood in terms of the rotation of magnetic domains along the field direction/ or the increase of volume fraction of the ferromagnetic domains.[29] The grain boundary contribution to MR (GBMR) of the compounds ($x \leq 0.05$) is estimated from back-extrapolating the linear high field MR and finding its intercept at zero field. GBMR is found to be ~ 23 % with no systematic variation with x (**Table 2**). On the other hand, for x = 0.07, MR shows almost linear dependence up to 5 T and similar to behaviour observed in the insulating cobaltate[30] showing cluster-glass behaviour. Additionally, field cycling leads to a non-closure of MR at zero field (**inset of the bottom panel of Fig.11**) indicating the presence of strong hysteresis and the magnetic after-effects, a characteristic feature of CG state.

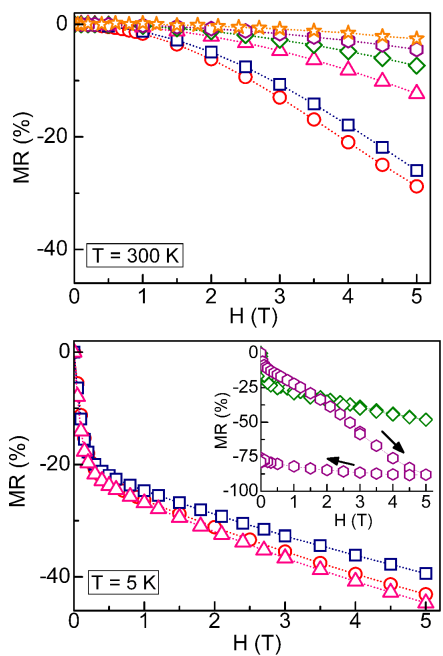

Fig.11 Field dependence of magnetoresistance (MR(H)) of $La_{0.67}Ca_{0.33}Mn_{1-x}Ti_xO_3$ ($0 \leq x \leq 0.10$) compounds at 300 K (top panel) and 5 K (bottom panel). Inset of the bottom panel shows the large hysteresis of MR of x = 0.07 with field cycling. The arrow indicates the direction of field sweep.

Yet another interesting feature to note is that MR(T) at 1 T for x = 0.07 shows a broad shoulder close to the corresponding $T_c$. Similar feature is also observed in Co substituted $La_{0.7}Ca_{0.3}MnO_3$[31] and the authors have argued that it is due to the formation of ferromagnetic clusters near $T_c$, which eventually freeze for $T \leq T_c$. Since the magnetic field broadens the maximum, similar feature was not distinct in the MR(T) curve at 5 T. Additionally, there is a sharp drop in MR signal just below the respective maximum (~ 45 K (24 K) for H = 1 T (5 T)) in the MR(T) curves. It is worth recalling that MR(H) isotherm at 5 K shows an almost linear dependence on H for H up to 5 T, similar to what has been observed for insulating cobaltates[30] showing cluster-glass (CG) behavior. Additionally, field cycling leads to a non-closure of MR at the zero field (**inset of the lower panel of Fig.11**) indicating the presence of strong hysteresis and the magnetic after-effects, a characteristic feature of CG state. Based on these observations and by comparing the $\chi'(T)$ of the substituted compounds showing similar feature,[32] it is inferred that Ti substitution modifies the magnetic ground state to a cluster-glass state.

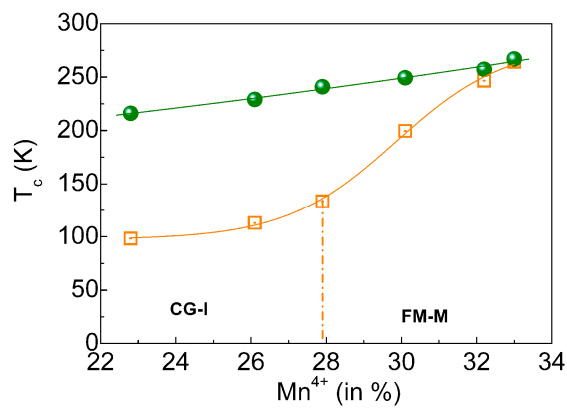

Fig. 12: Phase diagram of $La_{0.67}Ca_{0.33}Mn_{1-x}Ti_xO_3$ ($0 \leq x \leq 0.10$) compounds (□) as a function of relative concentration of $Mn^{4+}$ (%) observed. FM-M and CG-I denote the ferromagnetic-metal and cluster glass-insulator state of the substituted compounds. For an inter comparison, similar plot for $La_{1-z}Ca_zMnO_3$ compounds (●)(reproduced from S.-W.Cheong et.al. Colossal magnetoresistive oxides, Y. Tokura (Ed), Monographs in Condensed Matter Science, Gordon and Breach, London (2000)). with same $Mn^{4+}$ concentration as that of substituted compounds are also given. Solid line is only guide to eye.

In the following, the microscopic effects of Ti substitution (reduced $Mn^{4+}$ concentration and local structural effects) are discussed. The estimated $dT_c/dx$ (~ 26 K/at.%) of Ti substituted $La_{0.67}Ca_{0.33}MnO_3$ is much larger compared to that of other trivalent substitutions in $La_{0.67}Ca_{0.33}MnO_3$ compounds.[12, 14, 16, 17, 18] The XRD studies of the compounds of present study show that Ti substitution alters the average Mn valence towards 3+ to preserve the charge neutrality of the compounds. Such a shift is expected to change the charge carrier concentration (CCC). If CCC is a dominant factor affecting transport and magnetic properties, similar effect should also have been seen in $La_{1-z}Ca_zMnO_3$ compounds[33] with comparable $Mn^{4+}$ concentration as the $La_{0.67}Ca_{0.33}Mn_{1-x}Ti_xO_3$ ($0 \leq x \leq 0.10$) compounds. For an inter comparison, such curves are displayed in **Fig.12**. However, the Ti substituted compounds exhibit a much lower value of $T_c$. For instance, for x = 0.07, 26.1 % of the Mn ions are $Mn^{4+}$ leading to $T_c$ = 133 K and a low temperature insulating glassy behavior. On the other hand, ferromagnetism and metallicity is found in $La_{0.74}Ca_{0.26}MnO_3$ ($T_c \sim 215$ K)[33] with the same $Mn^{4+}$ concentration. Sahana et. al.[34] also observed that Ti substitution for $Mn^{4+}$ site decreases $T_c$ faster than decreasing the divalent ion concentration in the La site. For instance, $La_{0.74}Pb_{0.26}MnO_3$ showed a $T_c$ of 280 K. On the other hand, for $La_{0.6}Pb_{0.4}Mn_{0.91}Ti_{0.09}O_3$, $T_c$ is found to be 92 K. In both the samples, estimated $Mn^{4+}$ concentration through iodometric titration was $\approx$ 25 %.[34] Therefore, the effects of carrier density seems to be insufficient to account for the larger suppression in the transition temperatures observed in the present case. Hence it is expected that the larger structural modification involving a systematic linear increase in $<d_{Mn-O}>$ as well as a decrease in $<Mn-O-Mn>$ play a major role through decreasing the DE interaction strength.

In the present work, it is observed that Ti substitution results in CG state for $x \geq 0.07$. On the other hand, trivalent substitutions bring out CG behavior for much higher compositions.[12, 17, 18, 35] For instance, $Al^{3+}$ ion is shown to modify the ferromagnetic ground state of $La_{0.67}Ca_{0.33}Mn_{1-x}Al_xO_3$ to a CG at x = 0.15.[12] $Ga^{3+}$ is reported to show a CG state only for x = 0.25.[35] In contrast to this, $Fe^{3+}$ ion ($3d^5$) introduces CG state at 10 at. % of the substitution.[36] $Fe^{3+}$ ion, which is known to substitute $Mn^{3+}$ site with ionic size mismatch ($\Delta$IR) zero due to similar ionic radius, induces no structural effects.[37] However, the weakening of DE interaction strength and thereby the CG formation could be explained by the fact that $Fe^{3+}$ ion couple antiferromagnetically[38] with the neighboring Mn spins. The ability of diamagnetic tetravalent Ti to induce CG behavior at even lower concentration compared to that of $Fe^{3+}$ (paramagnetic), $Al^{3+}$ and $Ga^{3+}$ (both diamagnetic) substituted system shows that the local structural effects play a dominant role in these compounds.

**Summary**

Systematic increase in the unit cell parameters and a monotonic decrease in $Mn^{4+}$ with x in $La_{0.67}Ca_{0.33}Mn_{1-x}Ti_xO_3$ ($0 \leq x \leq 0.10$) indicate that $Ti^{4+}$ predominantly substitutes $Mn^{4+}$ ion. $T_{MI}$ and $T_c$ decrease at a rate of 26 K/at.% up to x = 0.05. Beyond that, $T_c$ shows a highly non-linear behavior and the ground state modifies to a cluster-glass insulator. The suppression in the transition temperatures is attributed to the weakening of the DE interaction strength due to local structural effects by increasing $<d_{Mn-O}>$ and decreasing $<Mn-O-Mn>$. Modification in the major carrier concentration seems to be insufficient to account for the strong reduction in the transition temperatures of the compounds. The inter-comparison of structural, transport and magnetic properties of $Al^{3+}$, $Ga^{3+}$, $Fe^{3+}$ and $Ti^{4+}$ substituted systems show that local structural effects play a decisive role compared to that of local spin coupling effects on the FM-M ground state of CMR manganites.

**Acknowledgement**

Authors thank Dr. T. Geethakumary, MSD, IGCAR, for providing the ac susceptibility set up. One of the authors, LSL, acknowledges Council of Scientific and Industrial Research, India for Senior Research Fellowship.